\documentclass[10pt,a4paper]{article}
\usepackage[utf8]{inputenc}
\usepackage{amsmath}
\usepackage{amsfonts}
\usepackage{amssymb}

\usepackage{ae,amssymb,amsmath,pstricks,epsfig,cite,makeidx}
\usepackage[margin=10pt,font=footnotesize,textfont=sf,labelfont=sf]{subcaption}
\usepackage[T1]{fontenc}

\usepackage{yfonts,color}
\usepackage{titlesec}

\titleformat*{\section}{\large\bfseries}
\titleformat*{\subsection}{\bfseries}
\titleformat*{\subsubsection}{\bfseries}
\titleformat*{\paragraph}{\bfseries}
\titleformat*{\subparagraph}{\bfseries}

\newcommand{\sinc}{{\rm sinc}}
\newcommand{\dif}{{\rm d}}
\newcommand{\abs}[1]{\left\vert#1\right\vert}

\newcommand{\qedsymbol}{\hspace{\fill}\rule{1.5ex}{1.5ex}}
\newtheorem{remark}{Remark}

\newtheorem{theorem}{Theorem}[section]

\newtheorem{defin}{Definition}[section]

\title{Eigenfunctions of Underspread Linear Communication Systems}
\date{October 14, 2015}
\author{Sergio Barbarossa and Mikhail Tsitsvero}

\begin{document}
\setcounter{footnote}{-1}

\maketitle

\markright{Eigenfunctions of Underspread Linear Communication Systems}

 \section[Eigenfunctions of Underspread Linear Communication
Systems\\(S.\ Barbarossa and M.\ Tsitsvero)]{Introduction}

\markright{Eigenfunctions of Underspread Linear Communication Systems}

 \footnotetext{Authors: {\bf Sergio Barbarossa} and {\bf Mikhail Tsitsvero}, Department of
Information Engineering, Electronics and Telecommunications,
University of Rome ``La Sapienza'', Via Eudossiana 18, 00184 Rome,
Italy (sergio.barbarossa@uniroma1.it, tsitsvero@gmail.com).}

\def\tbW{\mbox{\boldmath \tiny $W$}}
\def\tbR{\mbox{\boldmath \tiny $R$}}
\def\bGamma{\mbox{\boldmath $\Gamma$}}
\def\cbR{\mbox{\boldmath $\cal R$}}
\def\bLambda{\mbox{\boldmath $\Lambda$}}
\def\bphi{\mbox{\boldmath $\phi$}}
\def\bDelta{\mbox{\boldmath $\Delta$}}
\def\bsigma{\mbox{\boldmath $\sigma$}}
\def\cbR{\mbox{\boldmath $\cal R$}}
\def\bLambda{\mbox{\boldmath $\Lambda$}}
\def\bsigma{\mbox{\boldmath $\sigma$}}
\def\btau{\mbox{\boldmath $\tau$}}
\def\betha{\mbox{\boldmath $\eta$}}
\def\bA{\mbox{\boldmath $A$}}
\def\bB{\mbox{\boldmath $B$}}
\def\bC{\mbox{\boldmath $C$}}
\def\bR{\mbox{\boldmath $R$}}
\def\bF{\mbox{\boldmath $F$}}
\def\tbF{\tilde{\bF}}
\def\bPhi{\mbox{\boldmath $\Phi$}}
\def\bSigma{\mbox{\boldmath $\Sigma$}}
\def\bI{\mbox{\boldmath $I$}}
\def\bU{\mbox{\boldmath $U$}}
\def\bu{\mbox{\boldmath $u$}}
\def\bv{\mbox{\boldmath $v$}}
\def\bzeta{\mbox{\boldmath $\zeta$}}
\def\bV{\mbox{\boldmath $V$}}
\def\bh{\mbox{\boldmath $h$}}
\def\bu{\mbox{\boldmath $u$}}
\def\bw{\mbox{\boldmath $w$}}
\def\bs{\mbox{\boldmath $s$}}
\def\ber{\mbox{\boldmath $e$}}
\def\bff{\mbox{\boldmath $f$}}
\def\bg{\mbox{\boldmath $g$}}
\def\bd{\mbox{\boldmath $d$}}
\def\bx{\mbox{\boldmath $x$}}
\def\bX{\mbox{\boldmath $X$}}
\def\bY{\mbox{\boldmath $Y$}}
\def\bS{\mbox{\boldmath $S$}}
\def\bJ{\mbox{\boldmath $J$}}
\def\bxi{\mbox{\boldmath $\xi$}}
\def\bXi{\mbox{\boldmath $\Xi$}}
\def\bgamma{\mbox{\boldmath $\gamma$}}
\def\by{\mbox{\boldmath $y$}}
\def\bq{\mbox{\boldmath $q$}}
\def\bg{\mbox{\boldmath $g$}}
\def\bH{\mbox{\boldmath $H$}}
\def\tbH{\tilde{\bH}}
\def\hbH{\hat{\bH}}
\def\bI{\mbox{\boldmath $I$}}
\def\bc{\mbox{\boldmath $c$}}
\def\bG{\mbox{\boldmath $G$}}
\def\tbG{\tilde{\bG}}
\def\b0{\mbox{$\bf 0$}}
\def\bone{\mbox{$\bf 1$}}
\def\cbH{\mbox{\boldmath $\cal H$}}
\def\cbG{\mbox{\boldmath $\cal G$}}
\def\cbY{\mbox{\boldmath $\cal Y$}}
\def\cbF{\mbox{\boldmath $\cal F$}}
\def\cbS{\mbox{\boldmath $\cal S$}}
\def\bX{\mbox{\boldmath $X$}}
\def\bW{\mbox{\boldmath $W$}}
\def\bS{\mbox{\boldmath $S$}}
\def\cbU{\mbox{\boldmath $\cal U$}}
\def\bydef{:=}
\def\limN{{\lim_{N \rightarrow \infty}}}
\def\intinf{\int_{-\infty}^{\infty}}
\def\summ{\sum_{m=0}^M}
\def\sinc{\rm sinc}
\def\bydef{:=}

\index{system!underspread|(}
\index{underspread system!linear|(}
\index{underspread system!eigenfunctions of|(}
\yinipar{T}he knowledge of the eigenfunctions of a linear time-varying (LTV) system is a
fundamental issue from both theoretical and
applications points of view. Nonetheless, no analytic expressions are
available for the eigenfunctions of a general LTV system. 
However, approximate expressions have been proposed for slowly-varying operators. 
In particular, in \cite{matz-hlaw-98} it was shown that an underspread 
system can be well approximated by a normal operator, so that 
we can properly talk about eigendecomposition. The interesting result derived in 
\cite{matz-hlaw-98} is that the class of eigenfunctions of underspread
systems can be approximated by a set of signals obtained as shifted versions, in both time and frequency, 
of a given pulse waveform $g(t)$, which is well localized in the time-frequency domain. 
The validity of this approximation depends on the system spread along the delay and Doppler axes.
An alternative approach, for Hermitian slowly-varying operators, was proposed in \cite{siro-knig-80}, \cite{Sirovich-Knight-Asymptotic} and \cite{Sirovich-Knight-Exact} 
where the authors used the WKB (Wentzel-Kromers-Brillouin) method to derive a relationship between the
instantaneous frequency of the channel eigenfunctions and the
contour lines of the Wigner Transform of the operator kernel (or
Weyl\index{Weyl symbol} symbol).

In this paper, we will follow an approach similar to
\cite{siro-knig-80} and show that the eigenfunctions
can be found exactly for systems whose delay-Doppler spread function is
concentrated along a straight line  and they can be found in
approximate sense for systems having a spread function
maximally concentrated in regions of the Doppler-delay plane whose
area is smaller than one. The interesting results are that: i) the instantaneous frequency 
of the eigenfunctions is dictated by the contour level of the time-varying transfer function;
ii) the eigenvalues are restricted between the minimum and maximum value of the 
the system time-varying transfer functions, but not all values are possible, as the system
exhibits an inherent quantization.

\section{Channel Characterizations}

We will consider input/output relationship for a continuous-time LTV system $y(t)=(\mathbf{H}x)(t)$. This relationship may be represented as following\footnote{Limits of integration are assumed to be from $-\infty$ to $\infty$ if not explicitly specified}
\begin{equation}
\label{input_output}
y(t) = \left( \mathbf{H} x \right) (t) =\int h(t,\tau) x(t-\tau) \ \dif \tau = \int k(t,\tau) x(\tau) \ \dif \tau,
\end{equation}
where $h(t,\tau)$ is the channel response at time $t$ to an impulse sent at time $t-\tau$, while $k(t,\tau)$ is the channel response at time $t$ to an impulse sent at time $\tau$. We will refer to $h(t,\tau)$ as \textit{time-varying impulse response} and to $k(t,\tau)$ as \textit{time-varying impulse response kernel} \cite{bell-63}. Any linear time-varying channel may be equivalently characterized by the \textit{delay-Doppler spreading function} $S(\tau,\nu)$ 
\begin{equation}
\label{input_delay-Doppler-spread}
S(\tau, \nu) = \int h(t, \tau ) e^{-j2 \pi \nu t} \ \dif t
\end{equation}
or by the \textit{time-varying transfer function} $H(t,f)$
\begin{equation}
\label{input_output4}
H(t,f) = \int h(t, \tau ) e^{-j2 \pi f \tau} \ \dif \tau.
\end{equation}
%

The corresponding adjoint channel is denoted as $\mathbf{H^*}$ and it corresponds to the following input/output relation
\begin{equation}
\label{input_output5}
y(\tau) = (\mathbf{H^*} x) (\tau) = \int h^*(t,t-\tau) x(t) \ \dif t.
\end{equation}

\section{Channel Decomposition}
In general, LTV systems are not Hermitian (i.e. $\mathbf{H}\neq\mathbf{H}^*$), so that they may not admit the canonical
eigendecomposition over a set of orthonormal eigenfunctions with real-valued eigenvalues.  However, they may be properly
characterized by introducing the \textit{left and right singular functions} with corresponding \textit{real-valued singular values}. 

\begin{defin}
The functions $u_i(t)$, $v_i(t)$ are the \textit{left and right singular functions} of the system $\mathbf{H}$ with a corresponding \textit{singular value} $\sigma_i \in \mathbb{R}^+$ if the following relationships are satisfied
\begin{eqnarray}
\label{sing_functions1}
\sigma_i u_i(t) = (\mathbf{H}v_i)(t), \\
\label{sing_functions2}
\sigma_i v_i(t) = (\mathbf{H}^*u_i)(t).
\end{eqnarray}
\end{defin}
From these expressions it is clear that left and right singular functions are the eigenfunctions of the composite Hermitian systems $\mathbf{H}\mathbf{H}^*$ and $\mathbf{H}^*\mathbf{H}$, respectively, with corresponding eigenvalues equal to $\sigma_i^2$.

If the system impulse response is  square-integrable, i.e.
\begin{equation}
\label{square_integr}
\int \int |h(t,\tau)|^2 dt \dif \tau < \infty,
\end{equation}
then there exists \cite{gall-68} a sequence (finite or infinite) of positive decreasing numbers $\sigma_1\geq...\geq\sigma_i\geq...>0$, which are the singular values of this system, and two sets of corresponding orthonormal left and right singular functions $u_i(t)$ and $v_i(t)$. 

However, it is worth noting that there are at least a few important cases for which $h(t,\tau)$ is not square integrable,
as for example in  linear time-invariant (LTI) or linear frequency invariant (LFI) channels, i.e. when $h(t,\tau)$ is constant along $t$ or when $h(t,\tau)=m(t)\delta(\tau)$. 

\section{Exact and Approximate Solutions}

\subsection{Spreading Function Concentrated over a Straight Line}
Let us start considering a channel having spreading function $S(\nu,\tau)$ with support concentrated solely over a straight line on the $(\nu,\tau)$-plane, i.e.  $S(\nu, \tau) = g(\tau)\delta(\nu-\mu \tau - f_0)$. A practical example of this class could be a multipath channel with two paths. Using (\ref{input_delay-Doppler-spread}), the time-varying impulse response of such system is:
\begin{equation}
\label{imp_resp_line}
h(t,\tau) = g(\tau) e^{j 2 \pi \mu \tau t} e^{j 2 \pi f_0 t}.
\end{equation}
In this case, the singular functions can be expressed in closed form, as stated next.
\begin{theorem}
Let us consider a system $\mathbf{H}$ with spreading function $S(\nu,\tau) = g(\tau)\delta(\nu-\mu\tau-f_0)$, where $\mu$ and  $f_0$ are real parameters and $g(\tau)$ is a smooth function. Then the left and right singular functions and singular values of this system are
\begin{eqnarray}
\label{right_sf_line}
v_i(t) &= &e^{j 2 \pi f_i t} e^{j \pi \mu t^2}, \\
u_i(t) &= &e^{j 2 \pi f_0 t} e^{j \arg{K(f_i,\mu)}} v_i(t), \\
\sigma_i &= &|K(f_i,\mu)|,
\end{eqnarray}
where $K(f_i,\mu) = \int g(\tau) e^{-j 2 \pi f_i \tau} e^{j \pi \mu \tau^2} \, \dif \tau$.
\end{theorem}
The proof may be obtained by direct substitution of (\ref{imp_resp_line}) into (\ref{input_output}).

\begin{remark}
From (\ref{input_output4}), the time-varying transfer function of the system in (\ref{imp_resp_line}) is
\begin{equation}
\label{tr_fun_line}
H(t,f) = e^{j 2 \pi f_0 t} G(f-\mu t).
\end{equation}
At the same time, the instantaneous frequencies of the left and right singular functions are, respectively, 
$f^l(t)=f_i+\mu t$ and $f^r(t)=f_0+f_i+\mu t$.
Hence, combining this result with (\ref{tr_fun_line}), we can draw the conclusion that the instantaneous frequencies 
of the left and right singular functions correspond to regions in the time-frequency plane where the time-varying transfer function is constant.
\end{remark}

\subsection{Spreading Function with Limited Support}

Let us consider now the case where the spreading function is mainly concentrated over a limited support
in the delay-Doppler domain. If we consider the
composite systems $\mathbf{H}\mathbf{H}^*$, we can write
\begin{equation}
\label{composite_input_output}
\sigma_i^2 u_i(t) = \int \mathcal{K}(t,\theta) u_i(\theta) \ \dif \theta,
\end{equation}
where the kernel $\mathcal{K}(t,\theta)$ is defined as
\begin{equation}
\label{composite_kernel}
\mathcal{K}(t,\theta) = \int h(t, t-\tau) h^*(\theta, \theta - \tau) \ \dif \tau.
\end{equation}
In this case, it is not possible to express the singular functions in closed form. Nevertheless, 
using the WKB method, as shown in the Appendix, we can still provide an approximated 
closed form solution:
\begin{equation}
\label{solution_form_main}
u_i(t) = \sum_m A_{i,m}(t) e^{j \phi_{i,m} (t)},
\end{equation}
\begin{equation}
\label{curves_sigma3_main}
\abs{ H \left( t, \frac{\dot{\phi}_{i,m}(t)}{2 \pi} \right) }^2 = \sigma^2_i,
\end{equation}
\begin{equation}
\label{amplitude_wkb_main}
A_{i,m} (t) =  \left. \left[ \abs{ \frac{ \partial \abs{H(t,f)}^2} {\partial f} } \right]^{-1/2} \right| _ {f = \frac{ \dot{\phi}_{i,m}(t)}{2 \pi}}.
\end{equation}
\begin{remark}
The instantaneous frequency of each component of the left singular function is still given, as in the previous section, by the  curves in the time-frequency plane where the time-varying transfer function is constant. Furthermore, the instantaneous bandwidth of each component depends on the derivative of the  time-varying transfer function evaluated along the instantaneous frequency curves\footnote{The instantaneous bandwidth was defined in \cite{Cohen90} as $B(t)=|A'(t)/A(t)|$}.
\end{remark}

\begin{remark}
From (\ref{amplitude_wkb_main}), the amplitude $A_{i,m}(t)$ diverges at points where the partial derivative is zero. These points are denoted as \textbf{turning points}, in analogy with what appears in the WKB solution of Schroedinger equation in physics. In the neighbourhood of these points, the solution (\ref{curves_sigma3_main}) and (\ref{amplitude_wkb_main}) is no longer valid and the problem must be analyzed separately.
\end{remark}

\subsection{Area Rule}
\begin{figure}[!t]
\centering
\includegraphics[width=.6\textwidth]{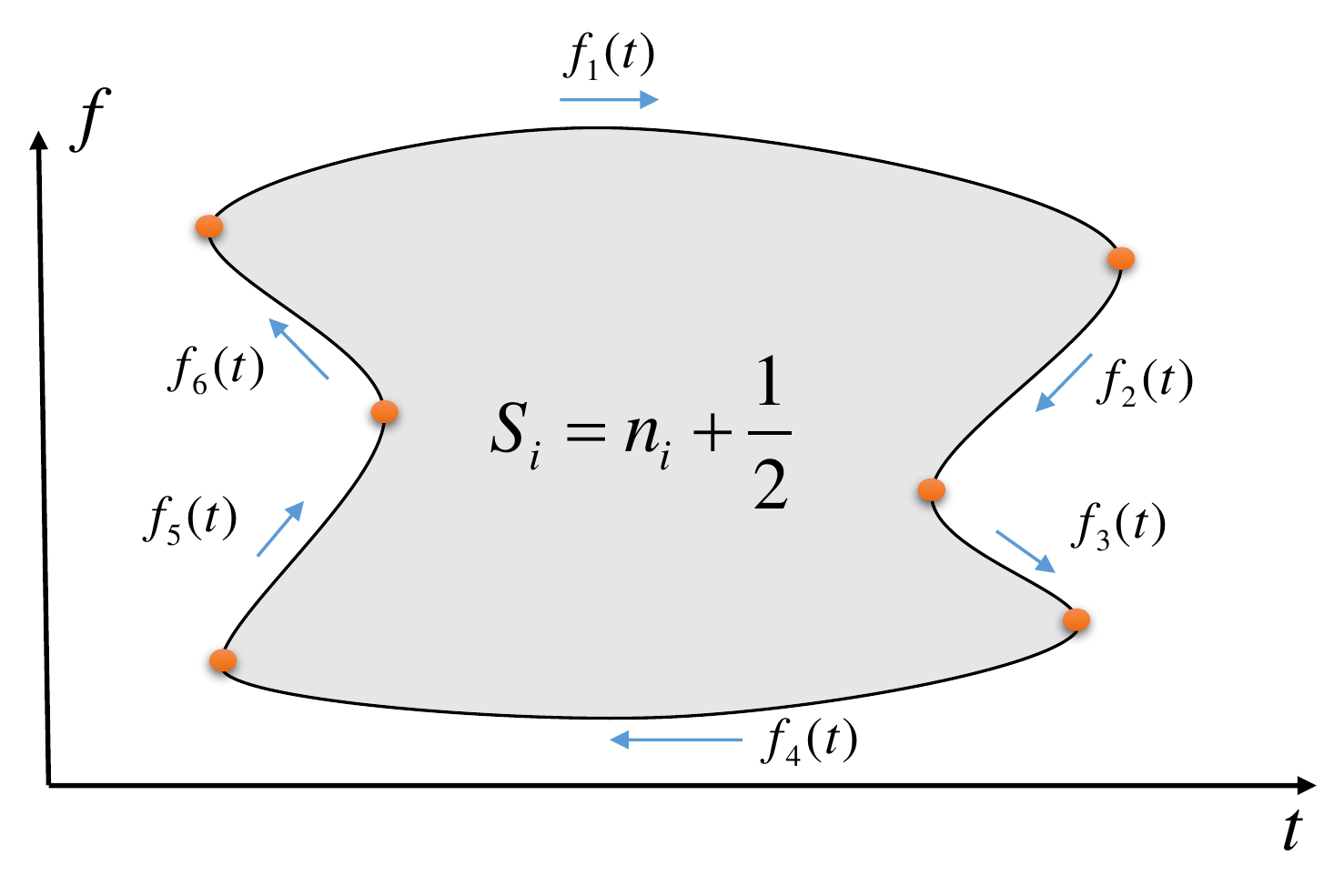}
\caption{Example of a bubble with six turning points}
\label{fig_area_rule}
\end{figure}
From (\ref{curves_sigma3_main}) it appears  that the singular values are bounded between the minimum and
maximum values of $|H(t, f)|$. However, it is still unclear if the singular values may assume any value within this interval. 
The answer is: no. It turns out that, 
under the square integrability assumption (\ref{square_integr}), the contour lines of $|H(t,f)|$ surface are always represented by a set of closed curves. In the following, we will refer to these closed curves as to \textit{bubbles} in time-frequency domain. A pictorial example of a bubble (contour plot of $|H(t, f)|$) is shown in Fig. \ref{fig_area_rule}.  Each bubble has an even number of turning points. Moreover, without any loss of generality, we will associate to each bubble its own eigenfunction $u_i(t)$ with a number of components  in the sum (\ref{solution_form_main}) equal to the number of turning points. What happens is that the only admissible values of $\sigma_i$ are the ones satisfying the so called \textit{area rule}, stating that the area of
the bubble associated to the level $\sigma_i$ must be equal to $n_i+1/2$, where $n_i$ is an integer nonnegative number.

This condition is known in physics as EBK (Einstein-Brillouin-Keller) quantization formula (see \cite{Sirovich-Knight-Asymptotic}). In our context, it gives a natural criterion for choosing the right values of $\sigma_i$. 
It is also worth to note here that the phase increments by $\pi/2$ at convex turning points and decrements by $\pi/2$ at concave turning points \cite{Sirovich-Knight-Asymptotic} for the traversing of the bubble as it is shown in Fig. \ref{fig_area_rule}.

\section{Numerical Validation}
In deriving analytic expressions for the singular functions we used approximations whose validity limits need to be checked.  Since the analytic models have a very specific structure in the time-frequency domain, the natural tool to analyze the results is a time-frequency representations with good time-frequency localization capabilities. To this purpose,  we use here the smoothed pseudo Wigner-Ville distribution (SPWVD) with a reassignment procedure, as developed by \cite{Flandrin-Reassignment} and \cite{Flandrin-tftb}. 
\subsection{Experimental Setup}
\begin{figure}[!t]
\centering
\includegraphics[width=\textwidth]{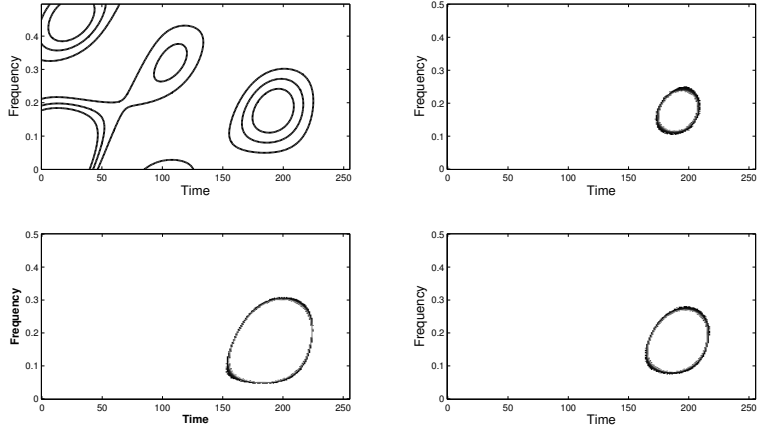}
\caption{Contour lines of $|H(t,f)|$ for the multipath case corresponding to three different levels (left top) and time-frequency distributions of singular vectors $\boldsymbol{u}_{25}$ (top right), $\boldsymbol{u}_{40}$ (bottom right) and $\boldsymbol{u}_{72}$ (bottom left)}
\label{fig_contour_comparison_multipath}
\end{figure}
We consider a channel with band limited input and time limited output, which may be approximately described as a windowing of time-varying transfer function $H(t,f)$ in both time and frequency. This leads to  a square-integrable time-varying impulse response $h(t,\tau)$.

By assuming input signal $x(t)$ to be bandlimited within the interval $[-1/2T_s, 1/2T_s]$ we can express it in accordance with the sampling theorem
\begin{equation}
\label{sampling_input}
x(t) = \sum_k x[k] \mathrm{sinc} \left( \frac{t-kT_s}{T_s} \right),
\end{equation}
where $\mathrm{sinc}(x) := \sin(\pi x)/\pi x$,  $x[k] := x(kT_s)$ and $T_s$ is a sampling time. By sampling the continuous time output $y(t)$ with same sampling rate $1/T_s$, we obtain the equivalent discrete-time input-output relationship
\begin{equation}
\label{discrete_io}
y[n] = \sum_k h[n,n-k] x[k],
\end{equation}
where $y[n] := y(nT_s)$ and $h[n,k]$ is the equivalent discrete time time-varying impulse response defined by
\begin{equation}
\label{discrete_impulse_response}
\begin{split}
h[n,n-k] &:= \int \int h(\theta, \tau) \mathrm{sinc} \left( \frac{nT_s - \theta}{T_s} \right)
 \mathrm{sinc} \left( \frac{\theta - \tau - kT_s}{T_s} \right) \dif \tau \dif \theta.
\end{split}
\end{equation}
Using (\ref{discrete_impulse_response}) the discrete-time counterpart of (\ref{input_output}) may be rewritten as 
\begin{equation}
\label{discrete_io_matrix}
\boldsymbol{y} = \boldsymbol{H} \boldsymbol{x},
\end{equation}
where $\boldsymbol{H}$ is the channel matrix. 
The left and right singular vectors $\boldsymbol{u}_i$, $\boldsymbol{v}_i$ of $\boldsymbol{H}$ can be
obtained by compting the singular value decomposition (SVD) of $\boldsymbol{H}$
\begin{equation}
\label{discrete_channel_matrix_decomposition}
\boldsymbol{H} = \boldsymbol{U} \boldsymbol{\Sigma}  \boldsymbol{V}^* ,
\end{equation}
where the columns of $\boldsymbol{U}$ and $\boldsymbol{V}$ are the left and right singular vectors $\boldsymbol{u}_i$, $\boldsymbol{v}_i$ associated to the singular value $\sigma_i$ contained in the diagonal matrix $\boldsymbol{ \Sigma }$.

We check the validity of our theoretical expressions through the following steps:
1) build the channel matrix $\boldsymbol{H}$; 
2) compute the SVD of $\boldsymbol{H}$;
3) compute the reassigned SPWVD (RSPWVD) of the singular vectors associated to some singular value $\sigma_i$, \cite{Flandrin-Reassignment};
4) compare the results obtained from (\ref{curves_sigma3_main}) and (\ref{amplitude_wkb_main}) with singular vectors computed numerically.

As we consider a windowed version of time-varying transfer function $H(t,f)$ the application of the area rule in some cases may not be straightforward because it is possible that the bubbles will be broken on the edges of the window. In these situations we observed an interesting feature that the area rule still holds exactly, however one should encounter just parts of the bubbles limited by the edges of the window.

\subsection{Multipath Case}
\begin{figure}[!t]
\centering
\includegraphics[width=.7\textwidth]{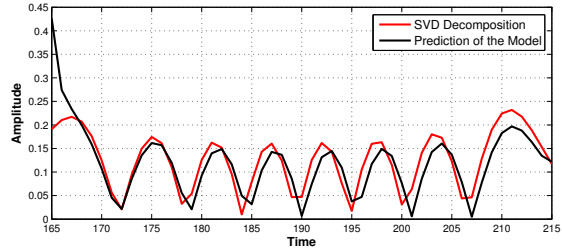}
\caption{Comparison of the amplitude of $\boldsymbol{u}_{40}$ with one, predicted by the model}
\label{fig_amplitude_comparison_multipath}
\end{figure}

As a test example, we consider a multipath channel with time-varying impulse response of the form
\begin{equation}
\label{multipath_imp_resp}
h(t, \tau ) = \sum_{q=0}^{Q-1} h_q e^{j 2 \pi f_q t} \delta(\tau - \tau_q),
\end{equation}
where each of the $Q$ paths is attenuated by $h_q$, delayed in time by $\tau_q$ and Doppler shifted by $f_q$. In accordance with Rayleigh fading model, we model $h_q$ as independent identically distributed (i.i.d.) complex Gaussian random variables having zero mean and unit variance. Time shifts $\tau_q$ and Doppler shifts $f_q$ are modeled as uniformly i.i.d. random variables within intervals $[0, \Delta \tau ]$ and $[-\Delta f/2, \Delta f/2]$ respectively.
In such a case, from (\ref{input_output4}), the time-varying transfer function is
\begin{equation}
\label{multipath_transfer_function}
H(t,f) = \sum_{q=0}^{Q-1} h_q e^{j 2 \pi f_q t} e^{ -j 2 \pi \tau_q f}.
\end{equation}
We provide the results of simulation for $Q = 10$ path channel, with $\Delta \tau = 4T_s$ and $\Delta f = 4/NT_s$, $N=256$. In Fig. \ref{fig_contour_comparison_multipath} three contour levels (corresponding to $\sigma_{25}$, $\sigma_{40}$ and $\sigma_{72}$) of $|H(t,f)|$ are given alongside with the time-frequency distributions (RSPWVD) of the corresponding singular vectors. We can observe  a very good agreement between the numerical results and the behaviors predicted by the proposed analytic models. Next we compare the amplitudes for numerically calculated singular vectors $\boldsymbol{u}_{40}$  with singular functions obtained by applying the formula (\ref{amplitude_wkb_main}). The result is shown in Fig. \ref{fig_amplitude_comparison_multipath}. One can note a slight departure between theoretical models and numerical results, essentially due to small errors around the turning points that accumulate in the computation of the 
instantaneous phase from its derivative using discrete integration methods (simple trapezium method).

\begin{figure}[!t]
\centering
\includegraphics[width=.6\textwidth]{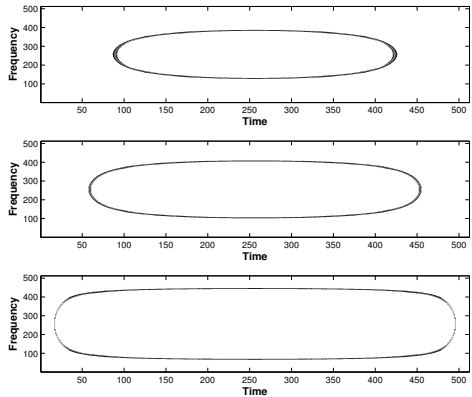}
\caption{Time-frequency distributions of prolate spheroidal functions}
\label{fig_prolate_spheroidal}
\end{figure}

As a further illustration of the nested structure of the bubbles in time-frequency domain, we consider the prolate spheroidal functions, known to be  the singular functions of the system approximately limited both in time and frequency (see \cite{Slepian1}, \cite{Landau1} and \cite{Landau2} for details). In Fig. \ref{fig_prolate_spheroidal} we report the contour plots of the RSPWVD of three prolate spheroidal sequences corresponding to three different singular values (decreasing - from top to bottom) of a given time and bandwidth limiting operators. It is interesting to observe how these waveforms fill the time-frequency region while being nested within each other rather than filling the same space through the usual rectangular tiling.

\section[Optimal Waveforms for LTV Digital
Communications]{Optimal Waveforms for Digital Communications
through LTV Channels}

\index{linear time-varying channel!optimal waveforms for|(}
One motivating application of the theory described above is digital communication through  
an LTV channel. In Article~13.2, for example, it is
shown how to convert the channel dispersiveness, possibly in both time
and frequency domains, into a useful source of diversity to be
exploited to enhance the SNR at the receiver. Here we show that if the
transmitter is able to predict the channel time-varying transfer%
\index{time-varying transfer function}
function, at least within the next time slot where it is going to
transmit, it is possible to optimize the transmission strategy and
take full advantage of the diversity offered by the channel
dispersiveness (see e.g.\ \cite{barb-scag-01} for more
details).

Considering a channel with approximately finite impulse response
of order $L$, we can parse the input sequence in consecutive
blocks of $K$ symbols and insert null guard intervals of length
$L$ between successive blocks to avoid inter-block interference.
If the symbol rate is $1/T_s$, the time necessary to transmit each
block is $K T_s$. For each $i$-th block, we must consider the
channel $h_i(t, \tau)$ obtained by windowing $h(t, \tau)$ in time,
in order to retain only the interval $[i K T_s, (i+1)K T_s]$, and
in frequency, keeping only the band $[-1/2 T_s, 1/2 T_s]$. The
optimal strategy for transmitting a set of symbols
$s_i[k]:=s[iK+k]$, $k=0, \ldots, K-1$, in the presence of additive
white Gaussian noise (AWGN), is to send the signal
\cite[sec.\,8]{gall-68} \vspace{-1mm}\begin{equation}x_i(t)=\sum_{k=0}^{K-1}c_{i, k}
s_i[k]v_{i, k}(t) \vspace{-1mm}\end{equation} where $v_{i, k}(t)$ is the right singular
function associated to the $k$-th eigenvalue of the channel
response $h_i(t, \tau)$ in the $i$-th transmit interval and $c_{i,
k}$ are coefficients used to allocate the available power among
the transmitted symbols according to some optimization criterion
\cite{barb-scag-01}. Using (\ref{sing_functions1}) and (\ref{sing_functions2}), the received
signal is thus \vspace{-1mm}\begin{equation}y_i(t)=\intinf h_i(t, \tau)
x_i(t-\tau)\ \dif \tau+w(t)=\sum_{k}c_{i, k} \sigma_{i, k}s_i[k]u_{i,
k}(t)+w(t), \vspace{-1mm}\end{equation} where $u_{i, k}(t)$ is the left singular function
associated to the $k$-th singular value%
\index{singular value}
 of $h_i(t, \tau)$ and
$w(t)$ is AWGN. Hence, by exploiting the orthonormality of the
functions  $u_{i, k}(t)$, the transmitted symbols can be estimated
by simply taking the scalar products of $y(t)$ with the left
singular functions, i.e.\ \vspace{-1mm}\begin{equation}{\hat s}_i[m]=\frac{1}{\sigma_{i, m}
c_{i, m}}\intinf y(t) u_{i, m}^*(t) \ \dif t= s_i[m]+w_i[m], \vspace{-1mm}\end{equation} where the
noise samples sequence $w_i[m]:=\intinf w(t) u_{i, m}^*(t) \ \dif t$
constitutes a sequence of i.i.d. Gaussian random variables. In this
way, the initial LTV channel, possibly dispersive in both time and
frequency domains, has been converted into a set of parallel
independent non-dispersive subchannels, with no intersymbol
interference, and the symbol-by-symbol decision is also the
maximum likelihood detector.%
\index{singular function|)}

Most current transmission schemes turn out to be simple examples
of the general framework illustrated in this paper. For example, in
communications through flat fading multiplicative channels,%
\index{channel!multiplicative}\index{multiplicative system}
 whose
eigenfunctions are Dirac pulses, hence the optimal strategy is time
division multiplexing. By duality, the optimal strategy for
linear time-invariant channels is orthogonal
frequency division multiplexing%
\index{orthogonal frequency division multiplexing (OFDM)}
 (OFDM). In the most general case (of underspread
channels), the optimal strategy would consist in sending symbols through
channel-dependent waveforms that fill the assigned time-frequency frame
according to the nested bubble-like structures illustrated in this paper.
\index{linear time-varying channel!optimal waveforms for|)}
Of course, the limit of validity of such an approach is dictated by the
capability to provide a sufficiently accurate short term prediction of the channel variation, 
based on past observations. 
\section{Summary and Conclusions}

The analytic model for the eigenfunctions of underspread linear
operators shown in this article, although approximate, shows that the
energy of the system eigenfunctions is mainly concentrated along
curves coinciding with level curves of the system transfer
function. Numerical results show that indeed the analytic model
fits very well with the numerical results. This model provides a general
framework for interpreting some current data transmission schemes and,
most importantly, gives a new perspective on the selection of the optimal waveforms 
to be used for transmissions over time-varying channels.


\section{Appendix}

To facilitate the analysis of slowly varying sysstems, the kernel $\mathcal{K}(t, \theta)$ may be rewritten in terms of the difference $t-\theta$ and the mean $(t+\theta)/2$ as:
\begin{equation}
\label{new_kernel}
\mathcal{K}(t,\theta) = K \left(t-\theta, \frac{\epsilon}{2}(t+\theta)\right),
\end{equation}
where $\epsilon \ll 1$ is a parameter representing the small dependence on $(t+\theta)/2$ (a time invariant system 
simply corresponds to $\epsilon=0$). This formulation suggests \cite{siro-knig-80}, \cite{Sirovich-Knight-Asymptotic} the following form for the solution of (\ref{composite_input_output})
\begin{equation}
\label{solution_form_wkb}
u_i(t; \epsilon) = \sum_k A_{i,k}(\epsilon t; \epsilon) e^{j \phi_{i,k} (\epsilon t; \epsilon ))/\epsilon}.
\end{equation}

To clarify the meaning of (\ref{new_kernel}) and (\ref{solution_form_wkb}) let us consider the following Taylor series expansion of the amplitude and phase in (\ref{solution_form_wkb}) around an arbitrary point $t = t_0$
\begin{equation}
\label{solution_form_expanded}
\begin{split}
u_i(t; \epsilon) = e^{j\sum_{n=0}^{\infty} (-1)^n \epsilon^{n-1} (t-t_0)^n \phi_{i}^{(n)}(\epsilon t_0) / n!} \\
\cdot \sum_{n=0}^{\infty}(-1)^n \epsilon^n A_{i}^{(n)}(\epsilon t_0) (t-t_0)^n/n!.
\end{split}
\end{equation}
From (\ref{solution_form_expanded}) it becomes clear that solution of the form (\ref{solution_form_wkb}) allows us to expand (\ref{solution_form_wkb}) in power series of $\epsilon$, by taking different orders for the phase and the amplitude.

Solutions of the form (\ref{solution_form_wkb}) are usually called WKB (Wentzel-Kromers-Brillouin) solutions and the associated method of finding approximate solutions is usually applied to differential operators, however \cite{Sirovich-Knight-Asymptotic} proved it to be very well applicable to the wide class of integral operators. What we consider here is just a particular case of the well developed perturbation theory. 

Here and thereafter we assume that all series expansions in $\epsilon$ converge, however it may not always be the case and it is subject to additional analysis and corrections. Now, if we put $\epsilon \approx 0$, we immediately come to the ``unperturbed'' solution in the form of approximate complex exponentials with constant amplitude. The ``unperturbed'' kernel (\ref{new_kernel}) will depend only on the difference $t-\theta$, therefore this case is generally equivalent to the case of LTI systems. On the other hand, $\epsilon = 1$ leads to the true solution of (\ref{composite_input_output}). So the solution is first examined by using power series of small parameter $\epsilon$ and then these results are extended to the solution of (\ref{composite_input_output}).

By making a change of variables $q = \epsilon t$ and $u = t - \theta$ in (\ref{new_kernel}) and considering for simplicity (and without loss of generality) only a single term in the sum of (\ref{solution_form_wkb}), we can rewrite (\ref{composite_input_output}) as
\begin{equation}
\label{composite_io_rewritten}
\begin{split}
\sigma_i^2 A_{i,m}(q; \epsilon) = \int K (u, q - \epsilon u / 2) A_{i,m} (q - \epsilon u; \epsilon ) 
\cdot e^{j \left( \phi_{i,m} ( q - \epsilon u) - \phi_{i,m} (q) \right)/ \epsilon} \ \dif u.
\end{split}
\end{equation}
By taking the limit $\epsilon \rightarrow 0$, we come to
\begin{equation}
\label{curves_sigma}
\int K(u, q) e^{- j u  \dot{\phi}_{i,m}(q)} \ \dif u = \sigma_i^2.
\end{equation}
Or, alternatively, by introducing the function $\tilde{K}(p,q)$
\begin{equation}
\label{definitionKp}
\tilde{K}(p, q) = \int K(u, q) e^{- j 2 \pi u p} \ \dif u,
\end{equation}
which is also referred to as Wigner transform of $K(u, q)$ \cite{Sirovich-Knight-Asymptotic}, we can rewrite (\ref{curves_sigma}) as
\begin{equation}
\label{curves_sigma2}
\tilde{K} \left( \frac{\dot{\phi}_{i,m}(q)} {2\pi}, q \right) = \sigma_i^2.
\end{equation}
This is an implicit equation showing that the instantaneous phase $\phi_{i,m}(q)$ must be chosen such that condition (\ref{curves_sigma2}) is satisfied. Operator $\mathbf{H}\mathbf{H}^*$ is hermitian by definition, therefore $\tilde{K}(p,q)$ is real.

Now after making Taylor series expansion of each term in (\ref{composite_io_rewritten}) around $\epsilon = 0$ we can write, taking into account (\ref{curves_sigma}) and considering only terms $\propto \epsilon$,
\begin{equation}
\label{expansion_for_amplitude}
\begin{split}
&\int \left[ \frac{u}{2} \frac{ \partial K(u,q)}{ \partial q} A_{i,m}(q) + uK(u,q) \frac{\dif A_{i,m}(q)}{\dif q} \right.  \\
&\left. - j A_{i,m}(q) K(u,q) \frac{u^2}{2}\frac{\dif^2 \phi_{i,m} \left( q \right)}{\dif q^2} \right] e^{-j u \dot{\phi}_{i,m}(q)} \ \dif u = 0.
\end{split}
\end{equation}
Equation (\ref{expansion_for_amplitude}) may be thought as condition to null the term proportional to $\epsilon$ in expansion of (\ref{composite_io_rewritten}).

For further simplicity of notation we introduce the functions $\tilde{K}_{i,j}(p,q)$ as
\begin{equation}
\label{derivatives_Kp_definition}
\tilde{K}_{i,j}(p,q) = \frac{\partial^{i+j} \tilde{K}(p,q) }{\partial p^i \partial q^j} .
\end{equation}

By noticing that 
\begin{equation}
\label{derivative_Kp}
\begin{split}
&j  \tilde{K}_{1,1}\left( \frac{\dot{\phi}_{i,m}(q)} {2\pi}, q\right) = 2 \pi \int u  \frac{\partial K \left( u,q \right)}{\partial q} e^{-j u \dot{\phi}_{i,m}(q)} \ \dif u \\
&- j 2 \pi \frac{\dif^2 \phi_{i,m}\left(q\right)}{\dif q^2} \int u^2 K( u, q ) e^{- j u \dot{\phi}_{i,m}(q)} \ \dif u
\end{split}
\end{equation}
we can restate (\ref{expansion_for_amplitude}) as
\begin{equation}
\label{solution_for_amplitude}
\frac{\dif}{\dif q} \left[ A_{i,m}^2 (q) \tilde{K}_{1,0}\left( \frac{\dot{\phi}_{i,m}(q)} {2\pi}, q \right) \right] = 0.
\end{equation}
This equation shows that the amplitude must be a solution of (\ref{solution_for_amplitude}).

As a next step we establish the relationship between function $\tilde{K}(p,q)$ and time-varying transfer function $H(t,f)$. It would be useful to represent $\tilde{K}(p,q)$ in terms of time-varying impulse response $h(t,\tau)$. Combining (\ref{composite_kernel}), (\ref{new_kernel}) and (\ref{definitionKp}) we can easily see that
\begin{equation}
\label{K_imp_resp}
\begin{split}
\tilde{K}\left( p, q \right) =  &\int \int h \left( \frac{u}{2} + \frac{q}{\epsilon}, \frac{u}{2} + \frac{q}{\epsilon} - \tau \right)
h^* \left( -\frac{u}{2} + \frac{q}{\epsilon}, -\frac{u}{2} + \frac{q}{\epsilon} - \tau \right) e^{- j u p} \ \dif u \dif \tau.
\end{split}
\end{equation}
And by exploiting the fact that $h(t,\tau)$ is the inverse Fourier transform of $H(t,f)$ with respect to $f$, we obtain
\begin{equation}
\label{K_transf_func}
\begin{split}
\tilde{K} \left( p, q \right) = &\int \int H \left( \frac{u}{2} + \frac{q}{\epsilon}, \nu \right)
 \cdot H^* \left( -\frac{u}{2} + \frac{q}{\epsilon}, \nu \right) e^{- j 2 \pi (p-\nu) u} \ \dif u \dif \nu.
\end{split}
\end{equation}

Following the spirit of the previous discussion we try to find the equation of the curve where $\tilde{K}(p,q)$ is constant. By denoting for simplicity $p(q) = \dot{\phi}_{i,m}(q) / 2 \pi$ and assuming that $H \left( \pm \frac{u}{2} + \frac{q}{\epsilon}, \nu \right) \approx H \left( \frac{q}{\epsilon} , \nu \right)$ we can rewrite $\tilde{K} \left( p(q), q \right)$ as
\begin{equation}
\label{K_transf_func_approx}
\begin{split}
\tilde{K} \left( p(q), q \right) & \approx \int \abs{ H \left( \frac{q}{\epsilon}, \nu \right) }^2 \int e^{- j 2 \pi (p(q) -\nu) u} \ \dif u \dif \nu 
 = \abs{ H \left( \frac{q}{\epsilon}, p(q) \right) }^2.
\end{split}
\end{equation}
Therefore, putting $\epsilon = 1$, we can write instead of (\ref{solution_form_wkb}), (\ref{curves_sigma2}) and (\ref{solution_for_amplitude})
\begin{equation}
\label{solution_form}
u_i(t) = \sum_m A_{i,m}(t) e^{j \phi_{i,m} (t)},
\end{equation}
\begin{equation}
\label{curves_sigma3}
\abs{ H \left( t, \frac{\dot{\phi}_{i,m}(t)}{2 \pi} \right) }^2 = \sigma^2_i,
\end{equation}
\begin{equation}
\label{amplitude_wkb}
A_{i,m} (t) =  \left. \left[ \abs{ \frac{ \partial \abs{H(t,f)}^2} {\partial f} } \right]^{-1/2} \right| _ {f = \frac{ \dot{\phi}_{i,m}(t)}{2 \pi}}.
\end{equation}

\end{document}